\documentclass[aps,prb,twocolumn,groupedaddress]{revtex4}

\usepackage{graphicx}
\usepackage{amsfonts}
\usepackage{amssymb}

\begin{document}


\title{Unconventional conductance plateau transitions in 
quantum Hall wires with 
spatially correlated disorder}

\author{Tohru Kawarabayashi and Yoshiyuki Ono}
\affiliation{Department of Physics, Toho University,
Miyama 2-2-1, Funabashi 274-8510, Japan}

\author{Tomi Ohtsuki}
\affiliation{Department of Physics, Sophia University,
Kioi-cho 7-1, Chiyoda-ku, Tokyo 102-8554, Japan}

\author{Stefan Kettemann}
\affiliation{I. Institut f\"{u}r Theoretische Physik, Universit\"{a}t Hamburg,
20355 Hamburg, Germany}

\author{Alexander Struck}
\affiliation{Department of Physics, University of Kaiserslautern, 
D-67663 Kaiserslautern, Germany}

\author{Bernhard Kramer}
\affiliation{International University Bremen, 28757 Bremen, Germany}

\date{\today}

\begin{abstract}
Quantum transport properties  in quantum Hall wires in the presence of 
spatially correlated random potential are investigated
numerically. 
It is found that the potential correlation reduces the localization length 
associated with the edge state,   
in contrast to the naive expectation that the potential 
correlation increases it.  
The effect appears as the sizable shift of quantized conductance 
plateaus in long wires, where the plateau transitions 
occur at energies much higher than the Landau band centers. 
The scale of the shift is 
of the order of the strength of the random potential and is 
insensitive to the strength of magnetic fields. 
Experimental implications are also discussed.
\end{abstract}


\maketitle

\section{Introduction}
Quantum transport property
of two-dimensional(2D) systems in strong magnetic fields 
has been one of the central issues of the condensed matter
physics.  
Much work has been performed 
in connection with the quantum Hall 
effect \cite{Ando,Aoki,PG,Hatsugai,Huckestein,KOK}. 
Recent discovery  of the quantum Hall effect 
in graphene \cite{graphene} also stimulates the theoretical interest.  
When the system is in a strong magnetic field, the so-called edge states are 
formed along the boundaries of the system \cite{Halperin,Buttiker}. 
The edge state 
corresponds to the classical skipping orbit along the boundary 
and is known to  be
less influenced by impurities and defects 
of the system \cite{MF}. It thus plays 
an important role in the quantum transport   
of a two-dimensional system in a strong magnetic field.

Numerical studies of the electron states of 
two-dimensional systems in a strong magnetic field in the 
presence of boundaries have already been 
performed by many authors \cite{SKM,OOSSC,OOK,Ando2,AA}. It has been 
shown that the edge state is well defined and is extended 
along the boundary as long as  its energy 
lies away from those of the bulk Landau subbands and the 
magnetic fields are strong enough. On the other hand, 
when its energy lies in the 
middle of the bulk Landau subbands, the edge state mixes with 
the bulk states by the impurity scattering, 
which leads to the localization of edge states 
\cite{Ando2,AA}. Recently, it has been demonstrated for 
the case of long quantum Hall wires with uncorrelated disorder potential 
that the conductance indeed vanishes when the Fermi energy 
is close to the centers of the bulk Landau subbands \cite{KKO,SKOK}.
The vanishing conductance can be understood as the consequence 
of the mixing between the bulk states and 
the edge states having opposite current directions at 
each end of the system, as has been confirmed by a quantitative 
comparison with analytical results. 
This transition between a quantized conductance and the insulator 
is called the chiral metal-insulator transition(CMIT) \cite{KKO,SKOK}.

In the present paper, the effect of potential correlation
in such systems is investigated numerically. 
The potential correlation is, in general,  important for the 
transport in low dimensions, since in some cases it 
changes the localization behavior drastically.
For a carbon nanotube, it has been shown that the 
potential correlation leads to the absence of back scattering 
\cite{AndoNakanishi}.
Even for pure 1D systems, dimer type correlation or 
long-range correlation of potential gives
rise to delocalized states \cite{DWP,ML}.  
For the two-dimensional 
bulk system in a magnetic field, 
the effect of potential correlation has also been 
studied based on the continuum model \cite{AndoUemura,OO2,Ando3} as well as 
the tight-binding model \cite{KS}. It has been  
demonstrated that 
the mixing between 
edge states is suppressed by the potential correlation \cite{OO2}. 
In the analysis on the critical states of the Landau bands, it is 
seen that the critical energies are 
insensitive to the potential correlation as long as the 
strength of the disorder is weak \cite{KS}.
It may then be expected naively that CMIT observed for the 
uncorrelated potential is suppressed and  
the quantized conductance steps are recovered when
the potential correlation is introduced.
We find indeed the suppression of CMIT which yields the 
recovery of the quantized conductance steps. Quite unexpectedly, however, 
it is found that the localization length associated with 
the edge states is suppressed by the potential correlation.
This causes the shift of the quantized conductance 
steps toward higher energies.  It is remarkable that the 
potential correlation suppresses the conduction of wires, since 
the potential correlation is usually expected to enhance the 
conduction of the system. 
We find that the scale of the shift is of the order of the 
strength of random potential and is insensitive to the 
strength of the magnetic field as long as the 
Landau levels are well defined. On the other hand, no shift appears 
in the square geometry. 
It is therefore necessary to analyze the effect of 
potential correlation on the edge states in long wires to understand 
these shifts of conductance plateaus.

\section{Model and Method}
We adopt the tight-binding 
model described by the following Hamiltonian on the 
square lattice  
\begin{equation}
 H = \sum_{<i,j>} V \exp({\rm i} \theta_{i,j}) c_i^{\dagger}
 c_j  + \sum_i \varepsilon_i c_i^{\dagger}c_i,
\end{equation}
where $c_i^{\dagger}(c_i)$ denotes the creation(annihilation) operator 
of an electron on the site $i$.  
The summation of the phases $\{ \theta_{i,j} \}$ 
around a plaquette is equal to $-2\pi\phi/\phi_0$,  
where $\phi$ is the magnetic flux through the plaquette 
and $\phi_0 = h/e$ stands for the flux quantum. The elementary charge and 
the Planck constant are denoted by $e$ and $h$, respectively.
All length-scales are
measured in units of the lattice spacing.
The site energies
$\{ \varepsilon_i \}$ are assumed to be distributed with  
the Gaussian probability density
\begin{equation}
 P( \varepsilon ) =  \frac{1}{\sqrt{2\pi\sigma^2}}
                     \exp (-\varepsilon^2/2\sigma^2) , \label{Gauss}
\end{equation}
and to have the spatial correlation as
\begin{equation}
 \langle \varepsilon_i \varepsilon_j \rangle = \langle \varepsilon^2 
 \rangle \exp (-|\mbox{\boldmath $R$}_i-
 \mbox{\boldmath $R$}_j|^2/4\eta^2). \label{corr1}
\end{equation}
Here $\sigma$ and $\eta$ are the parameters 
which determine the strength of the random 
potential and the strength of its spatial correlation, respectively. 
The position vector for the site $i$ is denoted by $\mbox{\boldmath $R$}_i$.
The spatially correlated potential is made from the uncorrelated 
potential $v_j$'s as \cite{KS}
\begin{equation}
 \varepsilon_i = \frac{\sum_j v_j \exp (-|\mbox{\boldmath $R$}_j-\mbox{\boldmath $R$}_i|^2/2\eta^2)}{\sqrt{\sum_j
 \exp (-|\mbox{\boldmath $R$}_j-\mbox{\boldmath $R$}_i|^2/\eta^2)}}.
 \label{sum}
\end{equation}
When the uncorrelated potential $v_j$'s obey the Gaussian distribution
with variance $\sigma$, 
it is easy to verify that $\varepsilon_i$ satisfies 
the relations in eqs.(\ref{Gauss}) and (\ref{corr1}). 
The parameter $\eta$, therefore, 
represents the range of the impurity potential.  
In the following, we specify the disorder
strength by a parameter $w=\sigma \sqrt{12}$, 
since it can be regarded as the effective width of the 
potential distribution.

We consider a system with the length $L$ and the width $M$. Two leads 
are attached to both ends of the system and the fixed boundary 
condition is assumed in the transverse direction. For the realization of 
the isotropic correlation for $\varepsilon_i$ 
in the sample region $L\times M$, we consider 
the additional regions of the width $5\eta$ outside of the sample 
region in performing 
the summation over $v_j$ in eq.(\ref{sum}).   

The two-terminal conductance $G$ is obtained by means of the 
Landauer formula
\begin{equation}
 G = (e^2/h) {\rm Tr }T^{\dagger}T,
\end{equation} 
where $T$ is the transmission matrix. We adopt the transfer 
matrix method\cite{PMR} to evaluate the transmission matrix.
Throughout the present analysis, 
we consider an independent impurity configuration 
for each value of 
energy $E$ and $\eta$.
The width $M$ is set to be 20. 
The smallest magnetic flux $\phi/\phi_0$ per plaquette
is $1/40$, and the corresponding magnetic length 
$l=\sqrt{\hbar/eB}$ is
about $2.5$, much smaller than the system width $M$. 
This leads to the existence of edge states 
along the boundaries.

\section{Numerical Results}
\begin{figure}
\includegraphics[scale=0.7]{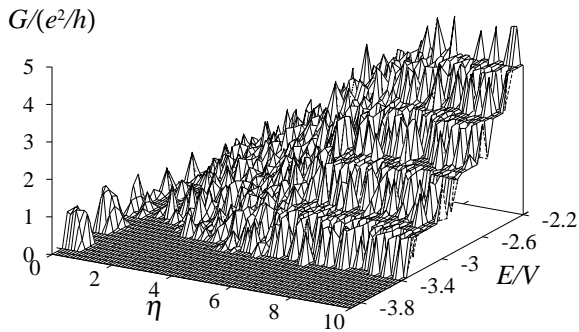}
\caption{Conductance for $w/V=0.8$, $\phi/\phi_0 = 1/40$ 
and $L/M=250$ as a function 
of the range of the impurity potential $\eta$ and the Fermi energy $E/V$.
In the present system, ratio of the with $M$ to the magnetic length $l$ is 
$M/l \approx 8$. 
\label{fig1}
}
\end{figure}

We first show the results of conductance for $w=0.8V$ and 
$\phi/\phi_0=1/40$ in Fig. \ref{fig1}. The length of the system is $L/M=250$.
In the absence of potential correlation ($\eta =0$), we see the 
chiral metal-insulator transition as a function of energies.
The conductance peaks at energies $E/V \approx -3.7$ and 
$-3.4$ are the contributions from the lowest and the 2nd
lowest Landau bands, respectively. It is clearly seen that 
these peaks disappear for $\eta > 1$ and that the quantized 
conductance steps are recovered for $\eta > 6$.
It is to be noted here that the recovery starts 
at $E/V \approx -3.4$. Since the contribution from 
the lowest Landau band starts at $E/V \approx -3.8$ for the case of 
$\eta =0$, the conductance steps can be understood as being shifted 
toward higher energies with an amount of $\delta E/V \approx 0.4$ 
in the presence of long potential correlation.

\begin{figure}
\includegraphics[scale=0.7]{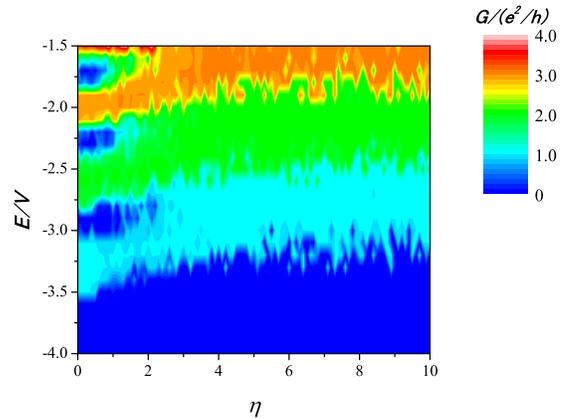}
\caption{(Color online) 
Conductance $G/(e^2/h)$ for $w/V=0.8$ and $L/M=50$ as a function 
of the potential range $\eta$ and the Fermi energy $E/V$.
The magnetic flux $\phi/\phi_0$ is assumed to be $1/16$. 
The magnetic length $l \approx 1.6$ and accordingly $M/l \approx 
12.5$.
\label{fig2}
}
\end{figure}

\begin{figure}
\includegraphics[scale=0.6]{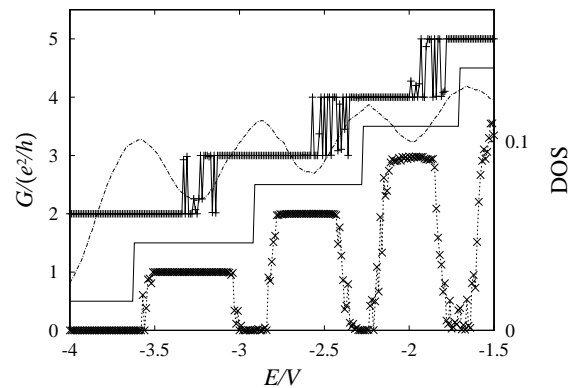}
\caption{Conductance $G/(e^2/h)$ 
for $\eta=0$ ($\times$) and $\eta =5 \approx 3.13l$ ($+$)
in the case of $w/V=0.8$, $\phi/\phi_0 = 1/16$ 
and $L/M=50$. For $\eta = 5$, $G/(e^2/h) + 2$ is 
plotted instead of $G/(e^2/h)$.  
Conductance $G_0$ for $w=0$ is also calculated and $G_0/(e^2/h) +0.5$
is plotted as a solid line. The density of states (DOS) for $\eta = 5$
obtained by the Green's function method \cite{SKM} 
is plotted as a dotted curve.
\label{fig3}
}
\end{figure}

In order to see the shift more clearly, we examine the conductance 
in the cases of stronger magnetic fields $\phi/\phi_0=1/16$ and $1/8$. 
In Fig. \ref{fig2}, the results 
for $\phi/\phi_0=1/16$ are shown. The length of systems 
is assumed to be $1000 (L/M=50)$. 
Here it is clearly seen that the conductance plateaus shift 
toward higher energies as the correlation length $\eta$ is 
increased. In both cases, the scale of the shifts in the 
presence of the long correlation turns out to be approximately $0.3V$
insensitive to the magnetic fields. 
We also evaluate the conductance for $L/M=1$, 
namely for the square system. It is then found
that no shift 
appears in that case. This is consistent with 
the result for the 2D bulk system that the levitation of critical 
states due to the potential correlation for the 
present range of disorder strength ($w/V \approx 1$) is 
much smaller than the present scale of the shifts of 
conductance plateaus \cite{KS}.
The present shift of conductance plateaus
is thus a specific feature to long wires, suggesting that
the shift arises from the one-dimensional character of the system 
rather than the 2D bulk properties. 
Apart from the shift of plateaus, it is also confirmed that 
CMIT fades away for larger values of $\eta ( \gtrsim 2.5)$.  
Detailed energy dependencies of the conductance in the presence of 
potential correlation ($\eta =5$) and that in its absence ($\eta=0$)
are shown in Fig. \ref{fig3}. It is clearly seen that the 
shifts are common to these three plateaus and CMIT 
observed for $\eta=0$ \cite{KKO,SKOK} disappears for $\eta=5$. 
The random jumps between the quantized plateaus for $\eta=5$ 
are fluctuations due to different realizations of disorder potential. 
It should be emphasized that the conductance 
plateau transitions occur away from the peaks of the density of 
states in the case of $\eta = 5$. 

In order to clarify how the scale of the shift depends on $w$,
we calculate the conductance also for $w/V=0.4$ and $1.6$
in the case of $L/M=50$ and $\phi/\phi_0 = 1/8$.
We obtain the shift to be $0.1\sim 0.2V$ and $0.6 \sim 0.7V$,
respectively. These results suggest that the scale of the 
shift is in proportion to $w$.

\begin{figure}
\includegraphics[scale=0.6]{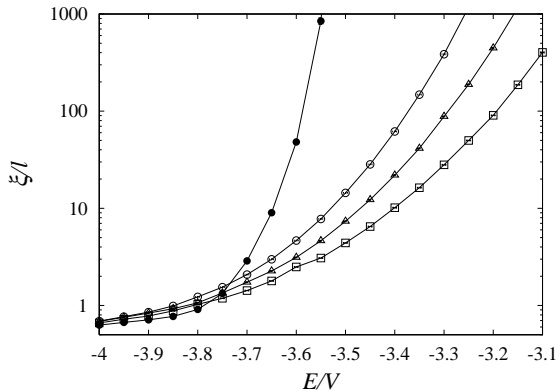}
\caption{Localization length along the wire of width 20  
in the cases of $\eta = 0(\bullet)$, 
$3(\circ )$, $5(\triangle )$ and 
$10(\square )$.
The disorder and the magnetic field are assumed to be $w/V=0.8$ 
and $\phi/\phi_0 =1/16$, respectively. 
The magnetic length is $l \approx 1.6$.
\label{fig4}
}
\end{figure}

The shift of plateaus means that the edge states which are 
transmitted in the case of $\eta =0$ are reflected for $\eta \neq 0$.
The localization length in the case of $\eta \neq 0$ must 
therefore be much smaller than that in the case of $\eta =0$. This 
is surprising since the potential correlation normally reduces 
the electron localization.
We thus examine the effect of potential correlation on the 
localization length along the wire in the presence of edge states.
The localization length $\xi$ estimated by the transfer 
matrix method \cite{MK} is shown in Fig. \ref{fig4} for 
$\phi/\phi_0=1/16$ and various values of $\eta$. Here we clearly 
see that the localization length $\xi$ is  much shorter 
than the system length $L/l=625$ for the energies $-3.6V \sim -3.3V$
in the presence of potential correlation.
This is consistent with the vanishing conductivity
in the presence of long potential correlation
for energies lower than $E/V \sim -3.3$  in Figs. \ref{fig2} and \ref{fig3}.

\section{Summary and Discussion}
Now we show 
the present shift of conductance plateaus  
can be understood as a semi-classical effect.
It is useful to recall here the fact that
the edge state is reflected by the potential 
barrier when its energy measured from 
the corresponding Landau level 
$\Delta E_{\rm edge}= E - (n+1/2)\hbar\omega_c$ 
is smaller than the energy of 
the potential barrier \cite{OO,Takagaki}, where $\omega_c$ denotes the 
cyclotron frequency. This suggests that the shift of conductance 
plateaus also occurs in the presence of potential barrier across the 
wire instead of disorder potential. We have confirmed in 
the present lattice model that the shifts expected in the continuum model
indeed occur when the thickness of 
the barrier is larger than the magnetic length.

With this property of the edge states, it is natural to expect 
that the edge states having 
lower energies ($\Delta E_{\rm edge} < w/2$) are deformed 
considerably by the correlated potential varying with scales larger 
than the magnetic length, because such a potential would 
act as a local potential barrier. 
Due to the deformation of the edge state trajectory,
in certain region of a long sample, the effective width of
the sample becomes narrow enough to induce the mixing of
edge states with opposite directions, leading to the
reflection.
As the potential correlation is 
increased, the deformation would  
become larger and the probability for the reflection of 
edge states is likely to increase, accordingly.
The same order of the shift ($\sim w/2$) 
in each plateau transition is also naturally understood since 
$\Delta E_{\rm edge}$ determines the probability of reflection.

The above semi-classical argument is consistent with our results 
that the shift becomes significant when the potential 
range $\eta/l$ is large enough
and that in the long correlation limit, it saturates around $w/2$ 
which is effectively the maximum of 
the potential energy (Fig. \ref{fig2}).
It would also account for the fact that the 
contributions from the lowest and the 2nd lowest Landau levels 
vanish when the potential range
$\eta$ approaches to the order of the magnetic length in Fig. \ref{fig1}.
In the presence of long potential correlation, it is thus expected that 
the conductance steps occur at critical energies $E_c \approx
(n+1/2)\hbar \omega_c + w/2$ in long wires. 
For certain fixed impurity configurations, we have
observed complex structure around the critical energy,
which is expected to be the resonant tunneling \cite{JK,RS} 
between edge states.

It is important to note that the mixing of the edge states at one 
place of the wire would be enough to reflect the whole current 
associated with them. 
The probability to have such a place in a particular sample 
apparently depends on the length of the wire.
It is natural that there is no shift in the case of the short system 
($L/M=1$) since such probability is very small. 
The probability $P$ to have high potential regions across the wire
in a sample is
estimated as $P \approx p^{M/\eta}(L/\eta)$ for a small 
$p$ which is 
the probability that the potential energy 
in the box of the size $\eta$ is larger than
$\Delta E_{\rm edge}$. 
Requiring $P$ is to be of order of unity, we get 
\begin{equation}
\log(L/\eta) \cong (-\ln p)(M/\eta), 
\end{equation}
which suggests that $M$ must be practically several times $\eta$ or less
to observe these phenomena.   
In reality, when the magnetic field is $10{\rm T}$, 
the magnetic length is $l\sim 8 {\rm nm}$, and the present systems 
correspond to wires whose width $M$ is in the range 
$70 \sim  140{\rm nm}$. The present numerical results suggest 
that the shift of conductance plateaus can be observed when  
$\eta >3l \sim 25{\rm nm}$ and $L= 3.5 \sim 7\mu {\rm m}$.

In summary, we have investigated the effect of potential correlation
in quantum Hall wires. It has been shown clearly that the 
potential correlation shifts the quantized conductance plateaus 
toward higher energies. The scale of the shifts is of the 
order of the strength of the random potential. 
This shift is specific to systems with the wire geometry and is 
insensitive to the strength of the magnetic fields.
This phenomenon is related to the 
transport property of edge states in the presence of long 
potential correlation. 
We have argued that the potential correlation enhances the 
mixing of the edge states at the opposite edges, which yields
the reflection of edge channels. 
For correlated potential, the positions of the 
conductance plateau transitions in quantum Hall wires 
do not necessarily coincide with the positions of the 
bulk Landau levels. The chiral metal-insulator 
transition is absent when the potential correlation is much larger 
than the magnetic length of the system.

\begin{acknowledgments}
This work was partly supported by a Grant-in-aid 
No. 16540294 and No. 18540382 
from the Japan Society for the Promotion of Science and by 
"Schwerpunktprogramm Quanten-Hall-Systeme", grant No. KE 807/2-1 
from the German research council (DFG). 
\end{acknowledgments}


\end{document}